\documentclass[twocolumn,twocolappendix]{aastex631}

\def\eg{{\it e.g.}}

\def\ie{{\it i.e.}}

\def\pmb#1{\setbox0=\hbox{$#1$}%
  \kern-0.25em\copy0\kern-\wd0
  \kern.05em\copy0\kern-\wd0
  \kern-0.025em\raise.0433em\box0}
\def\spmb#1{\setbox1=\hbox{${\scriptstyle #1}$}%
  \kern-0.25em\copy1\kern-\wd1
  \kern.05em\copy1\kern-\wd1
  \kern-0.025em\raise.0433em\box1}

\def\half{{\leavevmode\kern.1em\raise.5ex\hbox{\the\scriptfont0 1}\kern-.1em
    /\kern-.1em\lower.25ex\hbox{\the\scriptfont0 2}}}
\long\def\Ignore#1{\relax}

\def\spose#1{\hbox to 0pt{#1\hss}} 
\def\gtlt{\mathrel{\spose{\lower.5ex\hbox{$\mathchar"13E$}}
     \raise.5ex\hbox{$\mathchar"13C$}}}
\usepackage{color}
\definecolor{red}{rgb}{0.7,0.1,0.1}
\definecolor{blue}{rgb}{0.2,0.2,0.8}
\definecolor{green}{rgb}{0.1,0.6,0.1}

\begin{document}
\title{A Milky Way mass model for isolated simulations}
\shorttitle{A Milky Way model}

\author{J. A. Sellwood}
\affiliation{Steward Observatory, University of Arizona,
933 Cherry Avenue,
Tucson, AZ 85722, USA}
\email{sellwood@arizona.edu}

\shortauthors{Sellwood}

\begin{abstract}
  This paper presents an equilibrium model of a Milky Way-like spiral
  galaxy that supports open, mostly 2- and 3-arm spiral patterns but
  does not form a bar.  It is suggested as a more realistic
  alternative model to that employed by the {\sc agora} collaboration;
  their model has a much lower disk mass and therefore forms only
  multi-arm spiral patterns.  This improved model should enable
  simulations that test star-formation and feedback models in a more
  realistic isolated galaxy. Three versions of the same model having
  $2.1\times 10^5$, $2.1\times 10^6$, and $2.1\times 10^7$ particles
  are available for download.
\end{abstract}

\section{The AGORA model}
As an offshoot from the on-going {\sc agora} project
\citep[\eg][]{Kim14, Jung25}, \citet{Kim16} constructed an equilibrium
model of an isolated axisymmetric galaxy of roughly Milky Way mass.
Their purpose was to create a simplified model of an isolated galaxy
in which they could compare the performance of different codes, having
different recipes for star formation and feedback.  In order to ensure
that the evolution of their model supported only mild spirals and
avoided complications such as bar instabilities, they adopted a rather
low mass disk, as shown in Fig.~1; at its peak, the fraction of
central attraction from the disk, $V_d^2/V_{\rm tot}^2$, is about 1/3,
making it what is known \citep{Sack97} as a strongly submaximum disk.

We have downloaded the ``medium resolution'' version of their model
and evolved it using our {\sc galaxy} simulation code \citep{Sell14}.
The supplied particle file has 1M disk ``star'' particles, 1M ``gas''
particles, 0.125M bulge particles, and 1M halo particles.  The early
evolution of the star (top row) and gas (bottom row) disk is displayed
in Fig.~2, although in this simulation the ``gas particles'' were also
treated as collisionless.

As expected for a low-mass disk, the spirals are multi-armed
\citep{SM22}, and the disk does avoid forming a bar.  In the absence
of full gas physics, the low random velocities of the gas particles,
cause them to manifest filamentary spirals.

However, we here make a case that their model is not representative of
real galaxies.  We first note that more than 75\% of spiral galaxies
have 2- or 3-arm patterns \citep{Davis12, Hart16, YH18}, making the
evolution manifested in Fig.~2 quite untypical.  Furthermore,
multi-arm spiral patterns drive unrealistically slow dynamical
evolution of the disk.  The reasons for slow disk heating and radial
migration in a submaximal disk are set out in the Appendix.

\begin{figure}
\begin{center}
  \includegraphics[width=.8\hsize,angle=0]{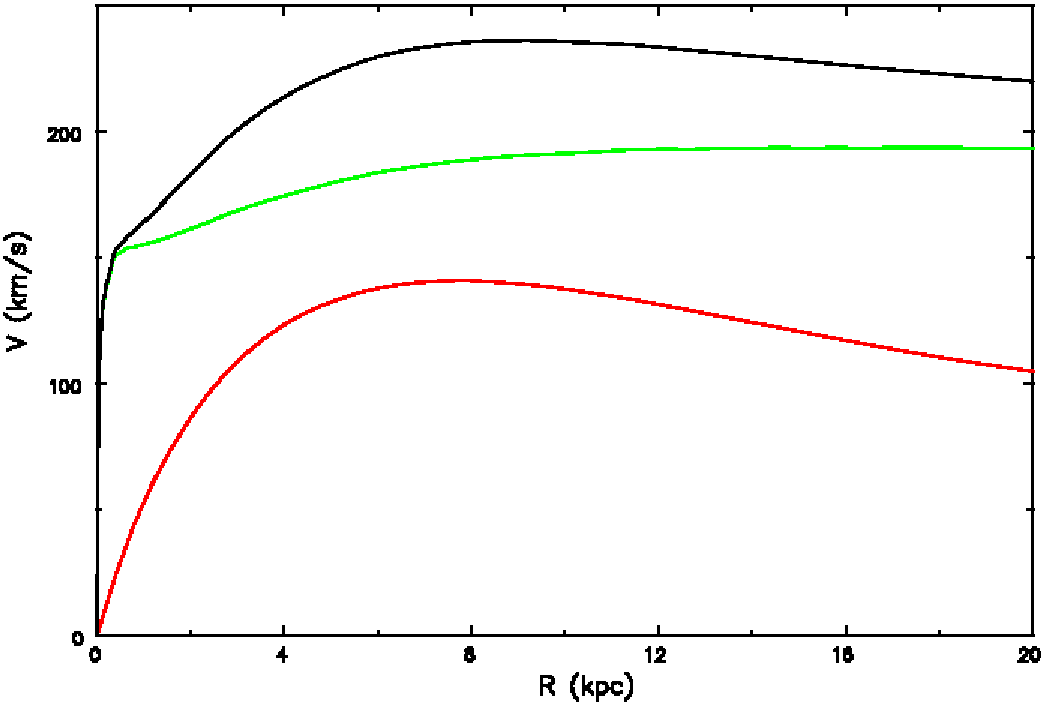}
\end{center}
\caption{The black curve shows the rotation curve of the {\sc agora}
  model. The red line shows the contribution from the disk star and
  ``gas'' particles and the green line the contribution from the bulge
  and halo.  It is evident that this model has a strongly submaximal
  disk.}
\label{fig.rc5191}
\end{figure}

\begin{figure*}
\begin{center}
  \includegraphics[height=.9\hsize,angle=270]{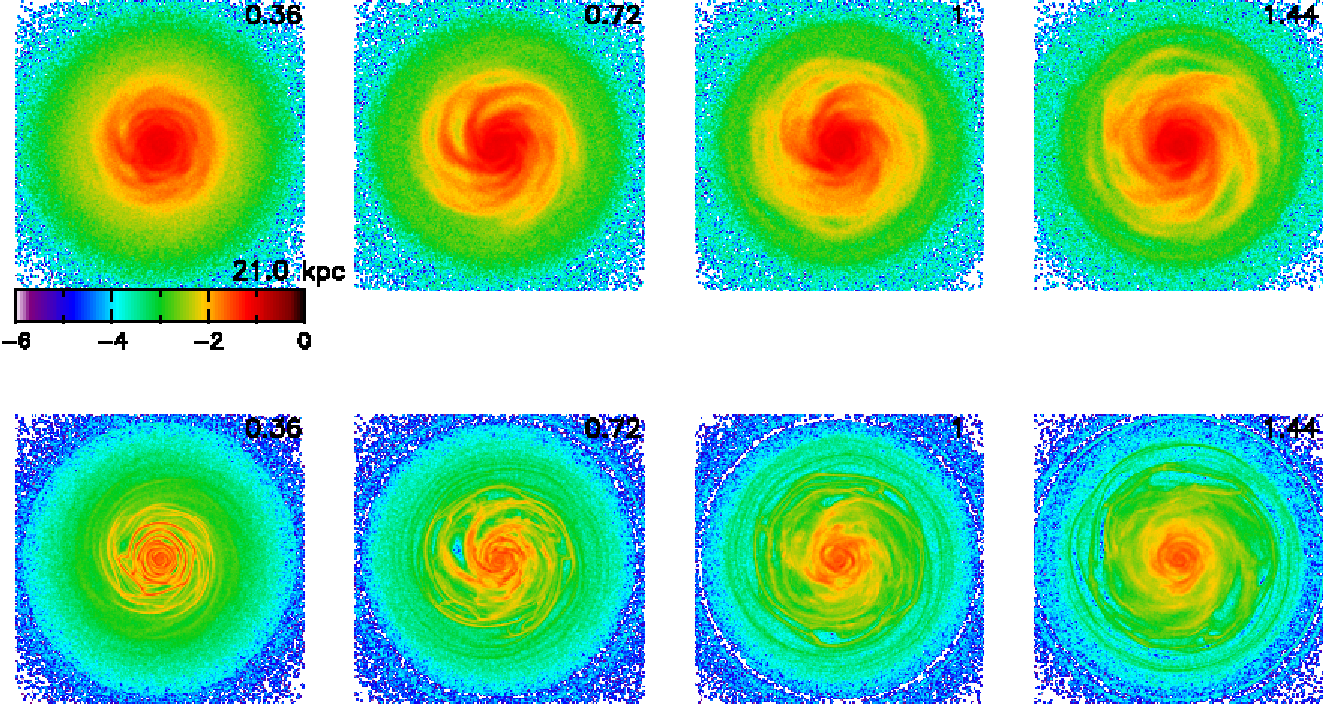}
\end{center}
\caption{The evolution of the disk particle density in the ``medium
  resolution'' {\sc agora} model.  The color scale indicates the
  logarithm of the disk surface density.  The top row shows the disk
  (star) particles, and the bottom the supplied gas particles,
  although here they are also treated as collisionless particles. The
  times are in Gyr and the box has a side of 42~kpc.  Note the
  multi-arm nature of the spiral patterns that develop in the stellar
  disk (top row).}
\label{fig.proj5937}
\end{figure*}

\section{Proposed alternative model}
Here we propose an alternative disk+bulge+halo model that differs from
the {\sc agora} model by having a more massive disk in order that it
will support fewer spiral arms, while the dense bulge component makes
the model less prone to bar instabilities.  The rotation curve of the
equilbrium model is shown by the black curve in Fig.~\ref{fig.MWRC}
and the separate contributions of the disk together with the bulge and
halo, which are all realized with collisionless particles, are also
shown.  A similar model, with a bulge of slightly lower central
density, was employed by \citet{RWSP}.

\subsection{Disk}
The stellar disk is a thickened exponential with radial scale length
$R_d$ having the volume density
\begin{equation}
\rho_d(R,z) = {M_d \over (2\pi)^{3/2} R_d^2}e^{-R/R_d} \exp\left(
    {-z\over 2z_0}\right)^2,
\label{eq.exp}
\end{equation}
with $z_0$ being the constant disk scale height.  We limit the radial
extent of the disk by tapering the density (eq.~\ref{eq.exp}) from its
value at $R=5.5R_d$ to zero at $R=6Rd$, which discards a few percent of
the mass.  We assign velocities to the disk particles as described in
\S\ref{sec.dvels} below.

\subsection{Bulge}
The spherical bulge has the cusped density profile proposed by
\citet{Hern90}
\begin{equation}
\rho(r) = {M_ba_b \over 2\pi r(r+a_b)^3},
\label{eq.bulge}
\end{equation}
with $M_b$ being the bulge mass and $a_b$ a scale radius.  Hernquist
also supplied an isotropic distribution function (DF) for this
isolated mass distribution, which we employ in this composite model to
set the bulge equilibrium, since the gravitational attraction in the
center is dominated by the bulge.

\subsection{Halo}
We sselect a second, more extensive and massive spherical Hernquist
model for the halo component, having mass $M_h$ and radial scale
$a_h$, and impose an outer boundary at $r=5a_h$ to the otherwise
infinite halo.  Since the central attraction of the disk and bulge,
which are embedded at the center of the halo, destroy the radial
balance of the isotropic DF supplied by Hernquist, we must derive a
revised equilibrium DF for the halo of the composite model.

Our method to achieve this is a follows: We start from the known
isotropic DF for the halo with no embedded disk or bulge and compute
its density change assuming that the masses of the disk and bulge were
increased adiabatically from zero.  As was pointed out by
\citet{Young80}, adiabatic changes to the total mass profile can be
calculated semi-analytically, since the actions \citep{BT08} of the
halo particles do not change as the potential well changes slowly;
therefore the DF expressed as a function of the actions is the same
after the adiabatic change as before.

\citet{Blum86} assumed angular momentum, one of the actions of an
orbit, was alone conserved, but their formula would apply only if the
orbits of all particles were initially and remained precisely
circular.  The orbits of particles in all reasonable spherical models
librate radially, and therefore one must take conservation of radial
action into account when computing the density response to adiabatic
changes to the potential, and the pressure of radial motions makes a
realistic halo more resistant to compression than a na\"{\i}ve model
with no radial action would predict.

\citet{Young80} and \citet{SM05} describe procedures, which we employ
here, to include radial action conservation, together with angular
momentum conservation, as the potential well changes adiabatically.
An initially isotropic DF becomes mildly radially biased, which can
still be represented by the unchanged actions.  Note that the
procedure assumes the potential remains spherically symmetric, so we
must approximate the disk potential by the monopole only term,
\ie\ the disk mass enclosed within a sphere of radius $r$.
\citet{SM05} found, from a comparison with a simulation in which a
flat disk was grown slowly inside a spherical halo that the aspherical
part of the disk potential caused negligibly small changes to the
spherically averaged halo potential.

\begin{figure}
\begin{center}
\includegraphics[width=.99\hsize,angle=0]{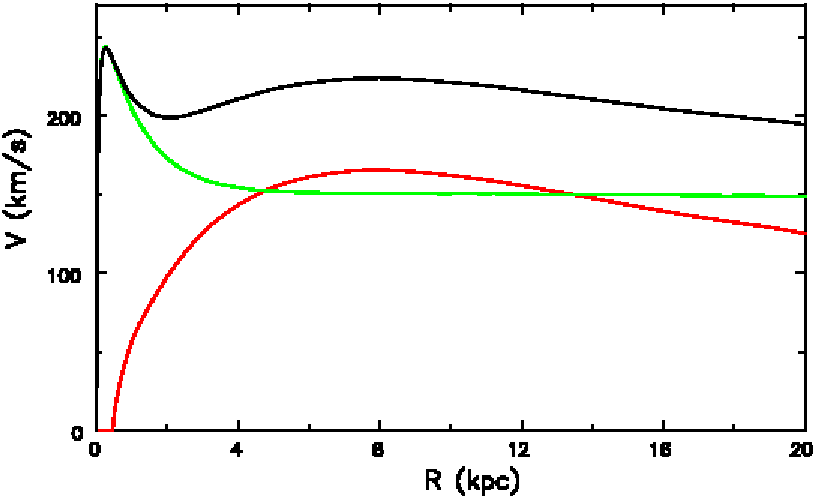}
\includegraphics[width=.99\hsize,angle=0]{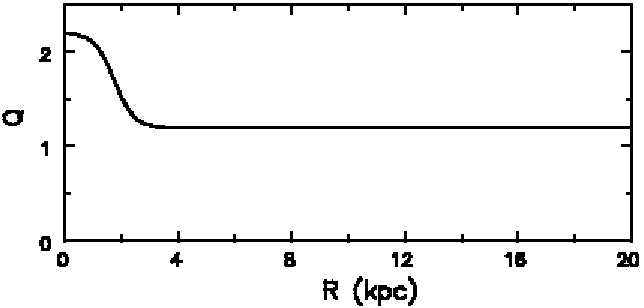}
\end{center}
\caption{Upper panel: The black line gives the total rotation curve of
  this proposed model, after halo compression.  The separate
  contributions of the thickened disk and the combined spherical bulge
  and halo are indicated by the red and green lines
  respectively. Lower panel: the initial radial dependence of Toomre's
  $Q$.}
\label{fig.MWRC}
\end{figure}

\subsection{Disk particle velocities}
\label{sec.dvels}
We use the grid to solve for the initial gravitational attraction of
the thickened disk (eq.~\ref{eq.exp}), which we combine with that of
the bulge and the now-known compressed halo density to obtain the
gravitational field everywhere.  We choose $Q=1.2$ \citep{Toom64} over
the outer disk, rising gently from $R=R_d$ to $Q=2.2$ at the center as
shown in the lower panel of Fig.~\ref{fig.MWRC} and select disk
particles from the DF proposed by \citet{Shu69}, with numerical
details given in \citet{Sell14}.  The vertical velocities are set from
the known density profile and potential, ignoring the small radial
gradients.

\begin{table}
\caption{Proposed Milky Way model}
\label{tab.pars}
\begin{tabular}{@{}lll}
  Disk & mass & $M_d = 6.10 \times 10^{10}\;$M$_\odot$  \\
            & scalelength & $R_d = 3.5\;$kpc  \\
            & vertical thickness & $z_0 = 350\;$pc \\
            & outer disk velocities & $Q=1.2$ \\
  Bulge &  mass & $M_b = 1.2 \times 10^{10}\;$M$_\odot$  \\
            & scale radius & $a_b = 245\;$pc  \\
            & velocity distribution & isotropic \\
  Initial halo & mass & $M_h = 1.2 \times 10^{11}\;$M$_\odot$  \\
            & scale radius & $a_h = 25\;$kpc  \\
            & velocity distribution & see text \\
\end{tabular}
\end{table}

\begin{figure*}
\begin{center}
\includegraphics[height=.8\hsize,angle=0]{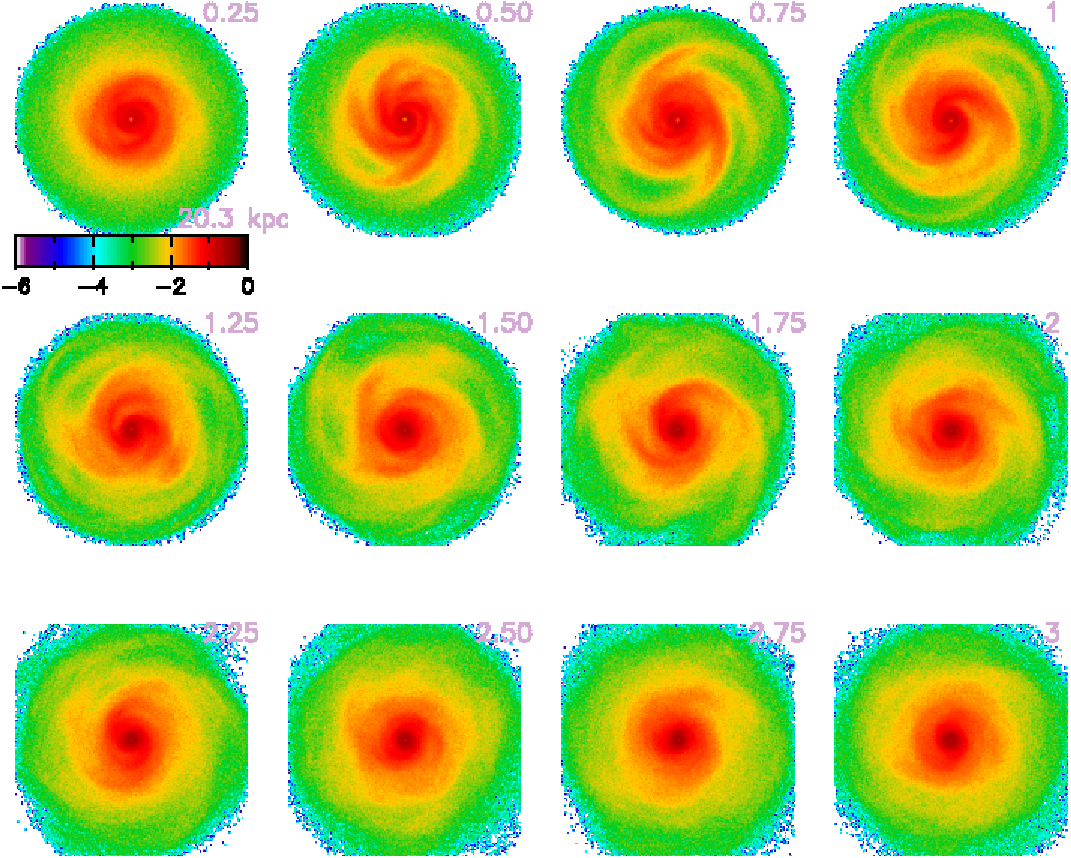}
\end{center}
\caption{The first 3~Gyr of evolution of our proposed model having no
  gas component.  The times are in Gyr and the half-widths of the
  frames are 20.3~kpc.  Notice the strong, open spirals and the
  absence of a large bar.}
\label{fig.proj}
\end{figure*}

\subsection{Selection of parameters and scaling}
Our choices for the parameters of the model are given in
Table~\ref{tab.pars} which, after halo compression result in the
rotation curve shown in Fig.~\ref{fig.MWRC}.  The disk is heavier than
that in the {\sc agora} model but at its peak $V_d^2/V_{\rm tot}^2
\simeq 0.58$, which is still not fully maximal \citep{Sack97}. We have
scaled the model such that the unit of time
$\left(R_d^3/GM_d\right)^{1/2} = 12.5\;$Myr.

\Ignore{
unit_L =    3.5000 kpc
unit_T =   12.5000 Myr
unit_V =  273.7896 km/s
unit_M =    6.1011 10^10 Msun}

We can realize this model with any number of particles in each
component, and find that it is reasonably close to equilibrium with
$T/|W| \simeq 0.515$.  Fig.~\ref{fig.proj} illustrates the first
3~Gyr of evolution of the disk, which is modeled by 1M star
particles.  There are no gas particles, but the bulge and halo (not
shown) are represented by 0.1M and 1M particles respectively.

The disk supports bi-symmetric and 3-arm spiral patterns but does not
form a bar.  The evolution of models having ten times more and ten
times fewer particles is quite similar.

\subsection{Including gas}
The model does not possess an explicit gas component.  Users planning
to implement gas physics based on particles, could simply relabel a
fraction of the collisionless disk partilces, selected at random (the
particles are ordered by binding energy), as gas and assign them gas
paticle properties.  Alternatively, for fluid-based gas phyics, the
mass of each disk particle can be reduced by the desired gas mass
fraction, while implementing the fluid code.  \eg\ In the simulations
reported by \citet{RWSP}, we reduced the mass of each disk particle by
10\% after 400Myr as the spiral patterns were beginning to develop,
and replaced this mass by a gas component modeled by the {\sc ramses}
code \citep{Teys02}.

\setbox1\hbox{\vbox to 2cm{{The website \filbreak
\noindent {\tt https://doi.org/10.25422/azu.data.30483479}
  \filbreak
  \noindent offers
a choice of three versions of the same model each having a different
number of particles:}}}

\section{Location and format of the files}
\medskip
\box1

\begin{itemize}
  \item Low resolution model: 0.1M disk particles, 0.1M halo particles
    and 0.01M bulge particles
  \item Medium resolution model: 1M disk particles, 1M halo particles
    and 0.1M bulge particles
  \item High resolution model: 10M disk particles, 10M halo particles
    and 1M bulge particles
\end{itemize}

Having selected the version you want, you can download the appropriate
gzipped tar ball and unpack it to obtain four files of ASCII data.
The disk, halo, and bulge files tabulate coordinates of every particle
in that component, one per line.  The first three numbers are the
$(x,y,z)$ position in units of kpc, the next three are the
$(v_x,v_y,v_z)$ velocity in units of km/s, and the final number is the
mass of the particle in units of $10^{10}$ solar masses.  All
particles within one component have the same mass, and the total mass
of the model (\ie\ the sum of the masses of all particles) is $\simeq
76 \times 10^{10}$ solar masses.
The last small file, named {\tt vcirc.dat}, is a table that gives the
circular speed (in km/s) at 1000 ordered radii (in kpc) to 50 kpc, and
is included for reference only.

I ask anyone wishing to use one or more of these models in a publication
to acknowedge their creator and cite this paper.

\newpage
\begin{acknowledgments}
I thank Steward Observatory for their continuing hospitality and the
Library of the University of Arizona for posting the files of
coordinates in a secure and globally accessible location.
\end{acknowledgments}

\appendix

The purpose of this appendix is to explain why multi-arm spirals heat
a submaximal disk more slowly and support less radial migration than
do patterns of greater scale in a heavier disk.
\begin{itemize}  
\item Spirals extract energy from the gravitational well of the galaxy
  by outward transport of angular momentum, and the energy adds to the
  random motion of the stars.  \citet{SB02} showed, without
  approximation or invoking any particular spiral arm theory, that the
  change of radial action, $\Delta J_R$, of stars for a given quantity
  of angular momentum, $\Delta L_z$, transported by a spiral pattern
  varies as
\begin{equation}
  \Delta J_R = \left|  \Delta L_z \right| / m,
\end{equation}
where $m$ is the angular periodicity, or number of arms in the
pattern.  This formula applies only at the Lindblad resonances where
the increase is positive at both the inner and outer LRs even though
the sign of the angular momentum change differs.  The factor $m$ in
the denominator is because the radial extent of a spiral pattern is
limited by its Lindblad resonances which are substantially closer to
corotation for multi-arm patters than for 2-arm spirals.  Therefore,
the efficiency of extraction of energy from the potential well by
larger $m$ spirals is reduced and the disk heats more slowly.

\item
  Mullti-arm patterns cause less radial migration \citep{SB02} for
  similar reasons.  To summarize the mechanism, a growing spiral
  perturbation traps stars onto horseshoe orbits at corotation.  As
  the amplitude rises, the width of the trapped region grows and
  trapped stars librate between the minima of the spiral potential
  where they reverse their motion in the rotating frame by gaining, or
  losing, angular momentum, which causes thenm to move onto a new near
  circular path at a different radius.  As the spiral decays, the
  trapped stars are released, and those that experienced an odd number
  (usually just one) of reversals while trapped will retain their new
  angular momenta, and therefore will have moved to orbits with
  greater, or smaller, guiding centers.

  The consequences for multi-arm spirals are:
\begin{enumerate}
  \item For a fixed spiral density amplitude, $\Sigma_a$ in eq.~(6.30)
    of \citet{BT08}, the perturbing potential is weaker for larger $k$
    (or shorter wavelength), which will diminish the radial extent of
    the horseshoe orbit region that is responsible for migration.
  \item The period of a horseshoe orbit trapped in the CR depends on
    both its frequency difference from CR and the azimuthal distance
    between wave-crests, which is shorter for higher-$m$ spirals.
    Since migration relies on the spiral having already decayed before
    a star makes its second horseshoe turn, those stars having periods
    long-enough to make only a single turn are confined to a narrower
    region about CR, implying a smaller average step size for
    migration.
  \item The increased value of $k$ causes the spiral potential to
    decay away from the mid-plane more rapidly, lessening its ability
    to affect the orbits of thick disk stars.
\end{enumerate}
These factors, which stem from the short wavelength of the spirals,
will reduce the extent of churning that is possible in both the thin
and thick disks in simulations of atypically sub-maximum disks, as
reported by \citet{Vera-C14}.  Note that \citet{Fran20} found strong
evidence for copious migration by stars in the Milky Way.
\end{itemize}

\bibliography{MW_model}{}
\bibliographystyle{aasjournal}

\end{document}